\documentclass[conference]{IEEEtran}
\IEEEoverridecommandlockouts
\usepackage{cite}
\usepackage{amsfonts}
\usepackage{algorithmic}
\usepackage{graphicx}
\usepackage{epstopdf}
\usepackage{textcomp}
\usepackage{url,,epsfig,bm,dsfont}%
\usepackage{amsmath} 
\usepackage{amssymb} 
\usepackage{subfigure}  
\usepackage{array}
\usepackage{subeqnarray}
\usepackage{cases}
\usepackage{enumerate}
\usepackage{lipsum}
\usepackage{multicol}
\usepackage{stfloats}
\usepackage{setspace}
\def\BibTeX{{\rm B\kern-.05em{\sc i\kern-.025em b}\kern-.08em
    T\kern-.1667em\lower.7ex\hbox{E}\kern-.125emX}}
\begin{document}

\title{The Effect of Mobility on Delayed Data Offloading\\
}

\author{\IEEEauthorblockN{Xiaoyi Zhou, Tong Ye}
\IEEEauthorblockA{\textit{School of Electronic and Electrical Engineering} \\
\textit{Shanghai Jiao Tong University}\\
Shanghai, China \\
zhouxiaoyi@sjtu.edu.cn, yetong@sjtu.edu.cn}
\and
\IEEEauthorblockN{Tony T. Lee}
\IEEEauthorblockA{\textit{School of Science and Engineering} \\
\textit{The Chinese University of Hong Kong, Shenzhen}\\
Guangdong, China \\
tonylee@cuhk.edu.cn}
}

\maketitle

\begin{abstract}
Delayed offloading is a widely accepted solution for mobile users to offload their traffic through Wi-Fi when they are moving in urban areas. However, delayed offloading enhances offloading efficiency at the expense of delay performance. Previous works mainly focus on the improvement of offloading efficiency while keeping delay performance in an acceptable region. In this paper, we study the impact of the user mobility on delayed data offloading in respect to the tradeoff between offloading efficiency and delay performance. We model a mobile terminal with delayed data offloading as an \emph{M/MMSP/1} queuing system with three service states. To be practical, we consider the feature of currently commercial mobile terminals in our analysis. Our analytical result shows that the mobility of the users can reduce the queueing delay incurred by the delayed offloading, and suggests that delayed offloading strategies can be optimized according to the mobility of the terminals once the delay requirement is given.
\end{abstract}


\section{Introduction}
In recent years, the urban areas have witnessed the surge in mobile data traffic. A Cisco’s survey shows that the global mobile traffic has grown 17-fold over the past 5 years \cite{CiscoWhitePaper}. The explosion of mobile traffic has led to the cellular network overloaded problem that causes the ruins of users’ satisfaction \cite{overload}. Even though 5G will bring new spectrum to market, but the proliferating of devices and connections will demand additional bandwidth \cite{wiaWhitePaper}. A widely accepted solution is to offload part of the mobile traffic through Wi-Fi interface with cheaper bandwidth \cite{pWhitePaper}. This solution currently becomes more and more attractive to both mobile network operators and mobile users.

Nowadays, cellular network coverage is nearly ubiquitous in urban areas. To provide offloading service, many public places, such as residential areas, commercial districts and transportation hubs, in urban areas are installed with Wi-Fi hotspots \cite{Wi-Fiplaces}. When the mobile users are moving in these places, they pass through Wi-Fi network coverages and cellular network coverages alternately. After losing Wi-Fi signal, if users can tolerate certain delay to wait for the next Wi-Fi access point, they will be able to offload more traffic through Wi-Fi, such that they can keep their communication costs as low as possible.

Based on such idea, several kinds of delayed offloading strategies have been proposed in \cite{buffer,howMuch,German,QoE,dynamicOffloading,802.11}. The goal is to promote offloading efficiency, while keeping delay performance in an acceptable region. Herein, the offloading efficiency is defined as the ratio of the data offloaded via Wi-Fi to the total transmitted data. In \cite{buffer}, service requests enter a Wi-Fi buffer when the buffer is not full; otherwise, the requests will be transmitted through the cellular network. In \cite{howMuch} and \cite{German}, each file, e.g., emails or pictures, is assigned with a timer when it enters the Wi-Fi buffer. If this file is still waiting in the Wi-Fi queue when the timer reaches a preset deadline, it will be sent via the cellular network. Ref. \cite{QoE} proposed to decide whether a newly arrived packet is directly transmitted via the cellular network, according to the Wi-Fi buffer length and the network connection. Clearly, these strategies implies that the mobile terminals are able to send the traffic through cellular network and Wi-Fi at the same time. However, such kind of concurrent transmission \cite{concurrentTransmission} is not supported by most of currently commercial mobile terminals \cite{multiFlow}.

In this paper, we study the data offloading problem of currently commercial mobile terminals, which can use only one kind of wireless channel to transmit traffic at the same time. Our goal is to find out if there is any factor that may affect the tradeoff between the offloading efficiency and the delay performance. We analyze the delayed data offloading by using an \emph{M/MMSP/1} queuing model with three service states. Our analytical results show that: though the offloading efficiency is enhanced at the expense of queuing delay, the moving speed of mobile users in motion is helpful to reduce the queuing delay incurred by the delayed offloading. This indicates that the deadline of delayed offloading strategies can be optimized according to the mobility of the terminals once the delay requirement is given.

The rest of this paper is organized as follows. In section II, we describe the delayed offloading strategy in detail, and show that the offloading procedure is essentially a three-state \emph{M/MMSP/1} queueing system. In section III, we establish a hybrid embedded Markov chain to derive the mean delay and the offloading efficiency. In section IV, we show that the moving speed of mobile users in motion can reduce the queuing delay incurred by the delayed offloading. Section V concludes this paper.

\section{Delayed Offloading of Terminals Without Concurrent Transmission}
When people are moving in the urban area, they pass through Wi-Fi coverages and cellular network coverages alternately. For a mobile terminal without concurrent transmission capability, it perceives the wireless channel switching between two states over time, as illustrated in Fig. 1, where $ C $ denotes the state that there is only cellular signal while $ F $ represents the state that the Wi-Fi signal is available.\par
Assume that the duration times of wireless channel states $F$ and $C$ are exponentially distributed with parameters $f_F$ and $f_C$, respectively. The wireless channel perceived by the mobile terminal in motion is a kind of Markov channel \cite{MarkovChannel}. Clearly, given the deployment of Wi-Fi hotspots, the faster the user moves, the larger $f_F$ and $f_C$ are. Thus, we use $f=\frac{1}{1/f_F+1/f_C}$ to delineate the mobility of the user.
\begin{figure}[b]
\centerline{\includegraphics[width=0.20\textwidth]{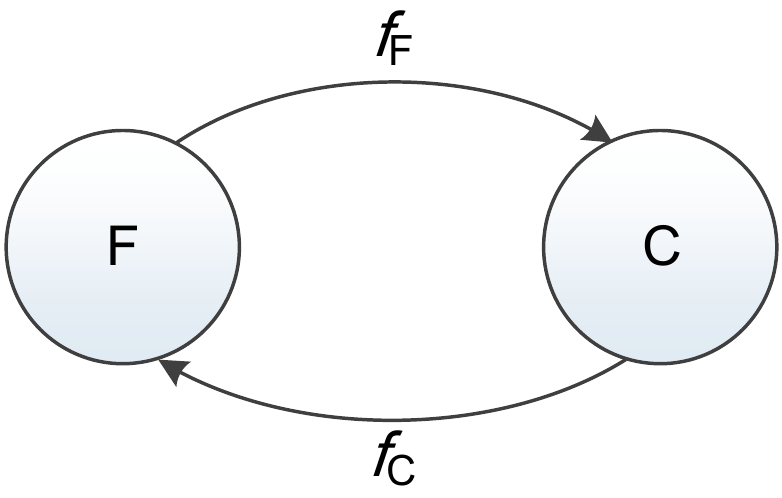}}
\caption{Transition of wireless environment state.}
\end{figure}
\subsection{Delayed Offloading Procedure}
In the face of such wireless environment, as described by the Markov channel in Fig. 1, we consider the delayed offloading procedure for currently commercial terminals as follows. When the Wi-Fi signal is available, the terminal transmits the traffic through Wi-Fi. Once the Wi-Fi connection is lost, the terminal pauses the traffic transmission to wait for the next Wi-Fi hotspot and randomly selects a deadline at the same time. If a Wi-Fi signal is available before the deadline expires, the terminal will recover the transmission through Wi-Fi; otherwise, it will go on with the transmission via the cellular network.

Hence, during the whole delayed offloading procedure, the terminal has three transmission (or service) states: (1) delayed state (or state 0), transmission is delayed, (2) cellular state (or state 1), transmission via the cellular network, and (3) Wi-Fi state (or state 2), transmission via Wi-Fi. The transitions among these service states are plotted in Fig. 2.

\subsection{Three-state Markov Modulated Service Process}
Suppose that the deadline set for the delayed state is an exponential random variable with parameter $f_D$. The data transmission process of mobile terminals can be considered as a three-state Markov modulated service process (MMSP) \cite{MMSP}. Let $f_{i,j}$ be the transition rate from state $i$ to state $j$ in Fig. 2, where $i$,$j=0,1,2$. According to Fig. 1 and 2, $f_{i,j}$ is given by
\begin{figure}[t]
\centerline{\includegraphics[width=0.26\textwidth]{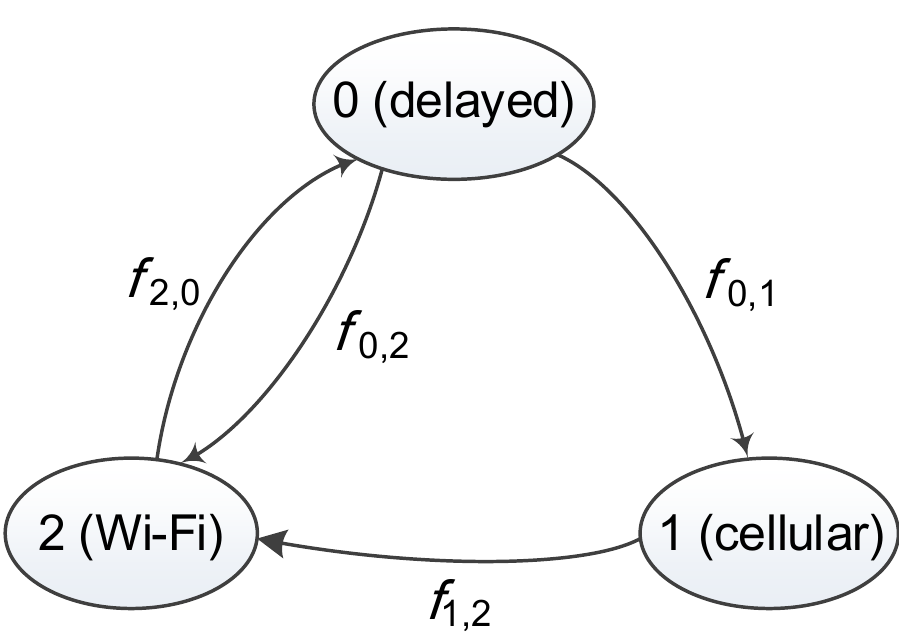}}
\caption{State transition of the data transmission.}
\end{figure}
\begin{subequations}
\begin{align}
&f_{2,0}=f_{F}\\
&f_{0,1}=f_{D}\\
&f_{0,2}=f_{1,2}=f_{C}.
\end{align}
\end{subequations}
It follows that the steady-state probabilities of each service state in Fig. 2 are given by
\begin{subequations}
\begin{align}
&\pi_0=\left(1\!-\!R\right)\frac{\tau f}{\tau f\!+\!1\!-\!R}\\
&\pi_1=\left(1\!-\!R\right)\frac{1\!-\!R}{\tau\! f\!+\!1\!-\!R}\\
&\pi_2=R,
\end{align}
\end{subequations}
where $\tau=1/f_D$ is the expectation of the deadline and $R=\frac{1/f_F}{1/f_C+1/f_F}=\frac{f_C}{f_C+f_F}$ is referred to as Wi-Fi available ratio in this paper since it actually indicates the ratio of the time that the terminal can perceive Wi-Fi signals.\par
Let $\mu_j$ be the transmission rate of state $j$ in Fig. 2. Clearly, the transmission rate of the delayed state is $\mu_0=0$. It follows that the average transmission rate that a mobile terminal with the delayed offloading strategy can provide is given by:
\begin{equation}
\hat{\mu}=\pi_1\mu_1+\pi_2\mu_2=\left(1\!-\!R\right)\frac{1\!-\!R}{\tau \!f\!+\!1\!-\!R}\mu_1\!+\!R\mu_2.
\end{equation}

\section{Analysis of Mean Delay and Offloading Efficiency}
In this section, we analyze the performance of the data offloading of the terminals without concurrent transmission capability. Suppose the input traffic is a Poisson process with rate $\lambda$. The data transmission of the offloading procedure can be delineated as an \emph{M/MMSP/1} queueing system with three service states. The difficulty of the analysis of the \emph{M/MMSP/1} queue lies in the fact that the service time of a file is related to the service state when its service starts. To cope with this problem, we use the hybrid embedded Markov chain developed in \cite{threeStates}.

\subsection{Embedded Points}
Two types of time points are embedded into the data offloading process. We consider the epoch when a file starts its service, since the service time of files depends on the service state at this epoch. We also observe the epoch at the transition of service states, since the dependency of the service time is essentially caused by service state transitions during the service of a file. We thus define the two types of embedded points as follows:
\begin{enumerate}[{1.}]
\item
State-transition point $\Phi_{j}$: epoch when the service state transits to state $j$;
\item
Start-service point $S_{j}$: epoch when a file starts its service while the service state is $j$.
\end{enumerate}
where $j=0,1,2$. Clearly, the time interval between two adjacent embedded points is exponentially distributed.

Suppose the current epoch is an embedded point of which the service state is the delayed state $j=0$, as Fig. 3(a) shows. Since the service is suspended at current epoch, the next event may be a state transition from service state 0 to service state $i$ after time $I_i$ which is an exponential random variable with parameter $f_{0,i}$, where $i=1,2$. Thus, the type of the next embedded point is determined by which kind of service state transition happens first. It follows that the distribution of the interval $I=\min\limits_{i}I_i$ from current point to the next point is exponentially distributed with parameter $\sum_{i=1}^{2}f_{0,i}$ and the next embedded point is $\Phi_i$ with probability $f_{0,i}/\sum_{i=1}^{2}f_{0,i}$, where $i=1,2$.\par
Similarly, when the current epoch is an embedded point of which the transmission state is state $j>0$, the next embedded point will be $\Phi_{\overline{j}}$ with probability $f_{j,\overline{j}}/\big(f_{j,\overline{j}}+\mu_{j}\big)$ or $S_j$ with probability $\mu_{j}/\big(f_{j,\overline{j}}+\mu_{j}\big)$, where $\overline{j}\triangleq\left(j+1\right)mod$ 3, as shown in Fig. 3(b) and (c). Also, the distribution of the interval from current point to the next point is exponentially distributed with parameter $f_{j,\overline{j}}+\mu_{j}$.
\begin{figure}[htbp]
\centering
\includegraphics[width=0.36\textwidth]{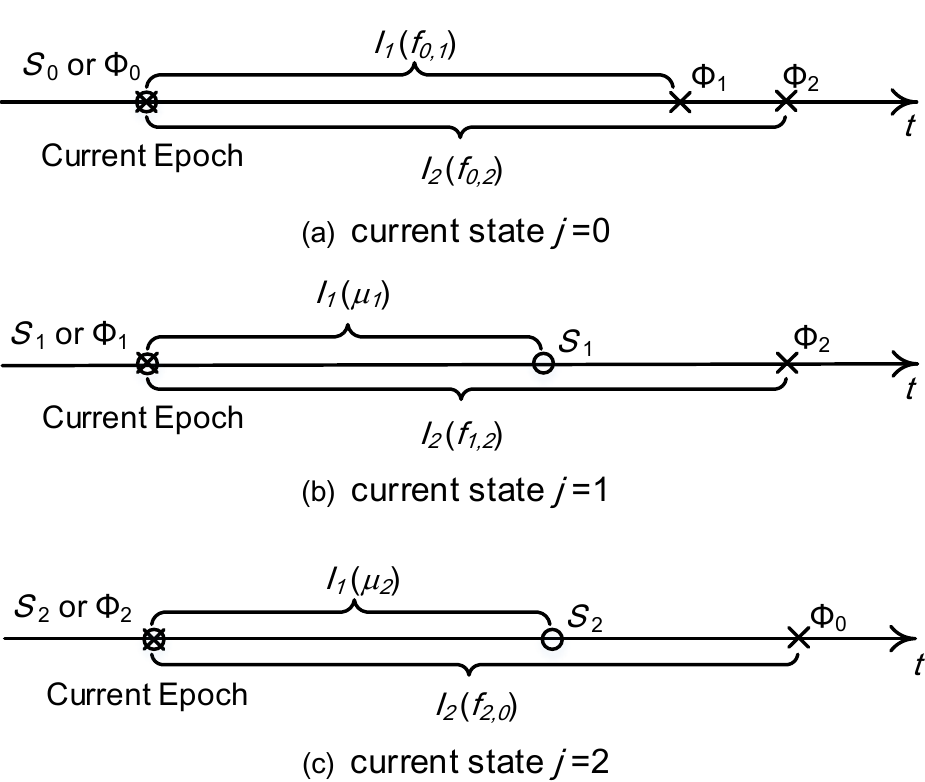}
\caption{Relationship between two kinds of embedded points.}
\end{figure}

\subsection{Start Service Probability}
The start service probability $\hat{\pi}_j$ is defined as the probability that a data file starts its service in state $j$. Consider a newly arrived file, which sees $n$ files in the buffer. These files are labeled according to their sequence in queue. The head-of-line (HOL) file is labeled with 0 and the newly arrived file is labeled with $n$. We define two types of conditional probabilities corresponding to the embedded points:

\newcounter{TempEqCnt}
\setcounter{TempEqCnt}{\value{equation}}
\setcounter{equation}{11}
\begin{figure*}[hb]
\begin{spacing}{2}
\hrulefill
\begin{subequations}
\begin{align}
&E\left[T_0\right]\!=\!\frac{\tau \!f^2\!+\!\left(1\!-\!R\right)f\!+\!\left(1\!-\!R\right)\tau f \mu_1\!+\!R\left(1\!-\!R\right)\left(\tau\! f\!+\!1\!-\!R\right)\!\mu_2\!+\!R\left(1\!-\!R\right)^2\!\tau \!\mu_1\mu_2}{\left(1\!-\!R\right)^2f\mu_1\!+\!R(\tau f\!+\!1-R)f\mu_2\!+\!R\left(1\!-\!R\right)\left(\tau \!f\!+\!1\!-\!R\right)\!\mu_1\mu_2}\\
&E\left[T_1\right]\!=\!\frac{\tau\! f^2\!+\!\left(1\!-\!R\right)f\!+\!R\left(1\!-\!R\right)\left(\tau\! f\!+\!1\!-\!R\right)\!\mu_2}{\left(1\!-\!R\right)^2f\!\mu_1\!+\!R(\tau f\!+\!1-R)f\mu_2\!+\!R\left(1\!-\!R\right)\left(\tau \!f\!+\!1\!-\!R\right)\!\mu_1\mu_2}\\
&E\left[T_2\right]\!=\!\frac{\tau \!f^2\!+\!\left(1\!-\!R\right)f\!+\!R\left(1\!-\!R\right)^2\!\mu_1\!+\!\left(1\!-\!R\right)\tau \!f \!\mu_1}{\left(1\!-\!R\right)^2\!f\!\mu_1\!+\!R(\tau f\!+\!1-R)f\mu_2\!+\!R\left(1\!-\!R\right)\left(\tau\! f\!+\!1\!-\!R\right)\!\mu_1\mu_2}
\end{align}
\end{subequations}
\end{spacing}
\end{figure*}
\setcounter{equation}{\value{TempEqCnt}}

\begin{enumerate}[{1.}]
\item
$\hat{\pi}_{n,j}\!\left(m\right)=P$\{the $m^{th}$ data file starts its service in service state $j$ $|$ the newly arrived file sees $n$ files in buffer\}
\item
$\hat{\varphi}_{n,j}\!\left(m\right)=P$\{the service state transits to state $j$ when the $m^{th}$ data file is in service $|$ the newly arrived file sees $n$� files in buffer\}.
\end{enumerate}
$\hat{\pi}_{n,j}\!\left(m\right)$ is defined on the embedded point $S_j$, at which the $(m-1)^{th}$ file finishes its service when the service state is $j$. Thus, the last event may be that the $(m-1)^{th}$ file starts its service when the service state is $j$ or that the service state transits to state $j$ when the $(m-1)^{th}$ file is in service. Therefore, for $1\le m\le n$, the equations of $\hat{\pi}_{n,j}\!\left(m\right)$ in each state are obtained:
\begin{subequations}
\begin{align}
&\hat{\pi}_{n,0}\!\left(m\right)=0\\
&\hat{\pi}_{n,1}\!\left(m\right)=\frac{\mu_1}{\mu_{1}\!+\!f_{1,2}}\Big(\hat{\pi}_{n,1}\!\left(m\!-\!1\right)\!+\!\hat{\varphi}_{n,1}\!\left(m\!-\!1\right)\Big)\\
&\hat{\pi}_{n,2}\!\left(m\right)=\frac{\mu_2}{\mu_2\!+\!f_{2,0}}\Big(\hat{\pi}_{n,2}\!\left(m\!-\!1\right)\!+\!\hat{\varphi}_{n,2}\!\left(m\!-\!1\right)\Big),
\end{align}
\end{subequations}
Similarly, the equations of $\hat{\varphi}_{n,j}\left(m\right)$ are given by:
\begin{subequations}
\begin{align}
&\hat{\varphi}_{n,0}\!\left(m\right)=\frac{f_{2,0}}{\mu_2+f_{2,0}}\Big(\hat{\pi}_{n,2}\!\left(m\right)\!+\!\hat{\varphi}_{n,2}\!\left(m\right)\Big)\\
&\hat{\varphi}_{n,1}\!\left(m\right)=\frac{f_{0,1}}{f_{0,1}+f_{0,2}}\Big(\hat{\pi}_{n,0}\!\left(m\right)\!+\!\hat{\varphi}_{n,0}\!\left(m\right)\Big)\\
\begin{split}
\hat{\varphi}_{n,2}\!\left(m\right)=\frac{f_{0,2}}{f_{0,1}+f_{0,2}}\Big(\hat{\pi}_{n,0}\!\left(m\right)\!+\!\hat{\varphi}_{n,0}\!\left(m\right)\Big)+\\
\frac{f_{1,2}}{\mu_1+f_{1,2}}\Big(\hat{\pi}_{n,1}\!\left(m\right)\!+\!\hat{\varphi}_{n,1}\!\left(m\right)\Big).
\end{split}
\end{align}
\end{subequations}
Combing (4) and (5), we have the relations between $\hat{\pi}_{n,j}\!\left(m\right)$ and $\hat{\pi}_{n,j}\!\left(m-1\right)$:
\begin{equation}
\begin{pmatrix}
  \hat{\pi}_{n,0}\!\left(m\right) \\
  \hat{\pi}_{n,1}\!\left(m\right) \\
  \hat{\pi}_{n,2}\!\left(m\right)
\end{pmatrix}
=\hat{Q}
\begin{pmatrix}
  \hat{\pi}_{n,0}\!\left(m-1\right) \\
  \hat{\pi}_{n,1}\!\left(m-1\right) \\
  \hat{\pi}_{n,2}\!\left(m-1\right)
\end{pmatrix}
=\hat{Q}^{m}
\begin{pmatrix}
  \hat{\pi}_{n,0}\!\left(0\right) \\
  \hat{\pi}_{n,1}\!\left(0\right) \\
  \hat{\pi}_{n,2}\!\left(0\right)
\end{pmatrix},
\end{equation}
where the coefficient matrix $\hat{Q}$ is
\begin{equation}
\begin{split}
&\hat{Q}=\\
&\begin{pmatrix}
\begin{smallmatrix}
  0 & 0 & 0 \\
  \beta \frac{f_{0,1}}{f_{0,1}\!+\!f_{0,2}}\!\left(1\!+\!\frac{f_{2,0}}{\mu_2}\right) & \beta \left(1\!+\!\frac{f_{0,1}}{f_{0,1}\!+\!f_{0,2}}\frac{f_{2,0}}{\mu_2}\right) & \beta\frac{f_{0,1}}{f_{0,1}\!+\!f_{0,2}}\!\frac{f_{2,0}}{\mu_2}\\
  \beta\left(\frac{f_{1,2}}{\mu_1}+\frac{f_{0,2}}{f_{0,1}\!+\!f_{0,2}}\right) & \beta\frac{f_{1,2}}{\mu_1} & \beta \left(1+\frac{f_{1,2}}{\mu_1}\right)
\end{smallmatrix}
\end{pmatrix},
\end{split}
\end{equation}
and
\begin{footnotesize}
\[
\beta\!=\!\frac{\!R\!\left(\!1\!-\!R\right)\!\left(\!1\!-\!R\!+\!\tau \!f\right)\!\mu_1\!\mu_2}{\left(1\!-\!R\right)\!^{2}\!f\!\mu_1\!+\!R\!\left(\!1\!-\!R\!+\!\tau\! f\right)\!f\!\mu_2\!+\!R\!\left(\!1\!-\!R\right)\!\left(\!1\!-\!R\!+\!\tau\! f\right)\!\mu_1\!\mu_2\!}.
\]
\end{footnotesize}
We solve (6) and obtain
\begin{subequations}
\begin{align}
&\hat{\pi}_{n,0}\!\left(m\right)=0\\
\begin{split}
\hat{\pi}_{n,1}\!\left(m\right)=\theta_1+\Big(\theta_2-\frac{f_{0,2}}{f_{0,1}+f_{0,2}}\Big)\beta^{m}\hat{\pi}_{n,0}\left(0\right)+\\
\theta_2\beta^{m}\hat{\pi}_{n,1}\left(0\right)-\theta_1\beta^{m}\hat{\pi}_{n,2}\left(0\right)
\end{split}\\
\begin{split}
\hat{\pi}_{n,2}\!\left(m\right)=\theta_2-\Big(\theta_2-\frac{f_{0,2}}{f_{0,1}+f_{0,2}}\Big)\beta^{m}\hat{\pi}_{n,0}\left(0\right)-\\
\theta_2\beta^{m}\hat{\pi}_{n,1}\left(0\right)+\theta_1\beta^{m}\hat{\pi}_{n,2}\left(0\right),
\end{split}
\end{align}
\end{subequations}
where $1\le m\le n$ and $\theta_{j}=\frac{\pi_j\mu_j}{\sum_{j=0}^{2}\pi_j\mu_j}$. When $m\!=\!0$, $\hat{\pi}_{n,j}\!\left(0\right)$ is the probability that the HOL file starts its service when the service state is $j$, given that the newly arrived file sees $n$ files in the buffer. Thus, $\hat{\pi}_{n,j}\!\left(0\right)=p_{n,j}/p_n$, where $p_{n,j}$ is the stationary probability that there are $n$ files in the buffer and the service state is $j$, and $p_n=\sum_{j=0}^{2}p_{n,j}$ is the stationary probability that there are $n$ files in the buffer. 

By definition, a newly arrived file that sees $n$ files in buffer upon its arrival starts its service in state $j$ is $\hat{\pi}_{n,j}\left(n\right)$. Thus, the start service probability $\hat{\pi}_j$ is
\begin{equation}
  \hat{\pi}_j=\sum_{n=0}^{\infty}p_n\hat{\pi}_{n,j}\left(n\right).
\end{equation}
Combining equations (8) and (9), we have:
\begin{subequations}
\begin{align}
\hat{\pi}_0=p_{0,0}\\
\begin{split}
\hat{\pi}_1=\theta_1+&\Big(\theta_2\!-\!\frac{\tau f}{\tau f+1-R}\Big)G_0\left(\beta\right)+\\
&\theta_2G_1\left(\beta\right)-\theta_1G_2\left(\beta\right)\!-\!\frac{1-R}{\tau \!f\!+\!1\!-\!R}p_{0,0}
\end{split}\\
\begin{split}
\hat{\pi}_2=\theta_2-&\Big(\theta_2\!-\!\frac{\tau f}{\tau f+1-R}\Big)G_0\left(\beta\right)-\\
&\theta_2G_1\left(\beta\right)+\theta_1G_2\left(\beta\right)\!-\!\frac{1-R}{\tau\! f\!+\!1\!-\!R}p_{0,0},
\end{split}
\end{align}
\end{subequations}
where $G_j\left(z\right)=\sum_{n=0}^{\infty}p_{n,j}z^n$. The numerical solutions of $G_j(z)$ and $p_{0,0}$ can be derived by establishing a two-dimensional continuous time Markov chain, which is not given in detail due to space limitation.

\subsection{Mean Service Time}
Let $T_j$ be the time needed to serve a file if the file starts its service in state $j$. Consider a file that the system is empty and in the delayed state when it arrives at the system. We say this file start its service in delayed state $j=0$. The service state changes to the state $i$ in the next embedded point with probability $f_{0,i}/\sum_{i=1}^{2}f_{0,i}$, where $i=1,2$. After that, the time this file still needed to finish the service is $T_i$. Considering that the time from current point to the next embedded point is $1/\sum_{i=1}^{2}f_{0,i}$, the expectation of $T_0$ is given by:
\begin{subequations}
\begin{equation}
\begin{split}
E\left[T_0\right]=\frac{f_{0,1}}{f_{0,1}\!+\!f_{0,2}}\left(\frac{1}{f_{0,1}\!+\!f_{0,2}}+E\left[T_1\right]\right)+\\
\frac{f_{0,2}}{f_{0,1}\!+\!f_{0,2}}\left(\frac{1}{f_{0,1}\!+\!f_{0,2}}+E\left[T_2\right]\right).
\end{split}
\end{equation}
Similarly, we obtain $E\left[T_1\right]$ in (11b) and $E\left[T_2\right]$ in (11c).
\begin{align}
  &E\left[T_1\right]\!=\!\frac{\mu_1}{\mu_1\!+\!f_{1,2}}\frac{1}{\mu_1\!+\!f_{1,2}}\!+\!\frac{f_{1,2}}{\mu_1\!+\!f_{1,2}}\!\left(\!\frac{1}{\mu_1\!+\!f_{1,2}}\!+\!E\left[T_2\right]\!\right)\!\\
  &E\left[T_2\right]\!=\!\frac{\mu_2}{\mu_2\!+\!f_{2,0}}\frac{1}{\mu_2\!+\!f_{2,0}}\!+\!\frac{f_{2,0}}{\mu_2\!+\!f_{2,0}}\!\left(\!\frac{1}{\mu_2\!+\!f_{2,0}}\!+\!E\left[T_0\right]\!\right)\!.
\end{align}
\end{subequations}
Solving (11), we can derive $E\left[T_j\right]$ in (12a)-(12c). And thus the mean service time:
\setcounter{equation}{12}
\begin{equation}
E\left[T\right]=\sum_{j=0}^{2}\hat{\pi}_jE\left[T_j\right].
\end{equation}

\subsection{Mean Waiting Time and Mean Delay}
The waiting time of a file is the duration that from the time it arrives at the system to the time it becomes the HOL file. It also equals to the sum of the residual service time of the HOL file and the service time of all the data files before this file. For the $k^{th}$ file in queue ($0\le k\le n$), we define two types of conditional elapse time:
\begin{enumerate}[{1.}]
  \item
  $W_{n,k}\!\left(k\right)$: the expected time from the epoch when a newly arrived file becomes the $k^{th}$ file in queue while the service state is $j$ to the epoch when it becomes the HOL file, given that it sees $n$ files in the buffer when it arrives;
  \item
  $V_{n,j}\!\left(k\right)$: the expected time from the epoch when the service state transits to state $j$ while the newly arrived file is now the $k^{th}$ file in queue to the epoch when it becomes the HOL file, given that it sees $n$ files in the buffer when it arrives.
\end{enumerate}
Following the similar arguments used to derive $\hat{\pi}_{n,j}\!\left(m\right)$, we have
\begin{equation}\label{iterative elapse time}
\begin{pmatrix}
  V_{n,0}\!\left(k\right) \\
  W_{n,1}\!\left(k\right) \\
  W_{n,2}\!\left(k\right)
\end{pmatrix}
=\sum_{i=1}^{k-1}\left(\hat{Q}^{T}\right)^i
\begin{pmatrix}
  E\left[T_0\right] \\
  E\left[T_1\right] \\
  E\left[T_2\right]
\end{pmatrix}
+
\begin{pmatrix}
  E\left[T_0\right] \\
  E\left[T_1\right] \\
  E\left[T_2\right]
\end{pmatrix}.
\end{equation}
Let $W$ be the mean waiting time. Solving (14) and using the relation $W=\!\sum_{j=0}^{2}\!\sum_{n=0}^{\infty}\!\left(n\right)p_{n,j}\!W\!_{n,j}$, we obtain:
\begin{equation}\label{mean waiting time}
\begin{split}
  W=\frac{1}{1-\frac{\lambda}{\hat{\mu}}}\bigg[\frac{\lambda}{\hat{\mu}}E\left[T\right]+&\frac{1}{1-\beta}\sum_{j=0}^{2}E\left[T\right]\left(\pi_j-\hat{\pi}_j\right)-\\
  &\frac{\beta}{1-\beta}\frac{\left(1-R\right)\tau}{1-R+\tau f}\left(\pi_0-\hat{\pi}_0\right)\bigg].
\end{split}
\end{equation}
Also, we have the mean delay as follows:
\begin{equation}\label{mean delay}
  D=W+E\left[T\right].
\end{equation}

\subsection{Offloading Efficiency}
Recall that the offloading efficiency, denoted by $\eta$, is defined as the ratio of the traffic transmitted via Wi-Fi to the total traffic. We consider a very long time period $\left[0,T\right]$. In this period, the total input traffic is $\lambda T$. On the other hand, $\pi_2-p_{0,2}$ is the probability that server is transmitting traffic while the service state is $j=2$, and thus the part of traffic served by Wi-Fi is $\left(\pi_2-p_{0,2}\right)\mu_2T$. It follows that
\begin{equation}
\eta=\frac{\mu_2}{\lambda}\left(\pi_2-p_{0,2}\right).
\end{equation}

\section{Performance Evaluations}
In reality, the deployment of Wi-Fi hotspots in a city is fixed. The factors that may affect the performance of the offloading procedure are the deadline set for the delayed service state and the moving speed of the users. Based on the analytical results in Section III, we study how the deadline and the moving speed of users affect the performances such as mean delay and offloading efficiency.

We consider two application scenarios in this section. One is that the terminals are carried by pedestrians, of which the channel transition rate $f_C=0.007s^{-1}$, $f_F=0.016s^{-1}$, and thus $f=0.005s^{-1}$. The other one is that the terminals are carried by vehicles. In this case, the channel transition rate $f_C=0.035s^{-1}$, $f_F=0.079s^{-1}$, and thus $f=0.025s^{-1}$ \cite{German}.

\subsection{$\mu_1=\mu_2=\mu$}
To facilitate discussion, we first consider the case where the transmission rate of the Wi-Fi connection and the cellular network are the same $\mu_1=\mu_2=\mu$.

We plot the offloading efficiency $\eta$ and the mean delay $D$ versus the expectation of the deadline $\tau$ in Fig. 4, where $\lambda=0.1$ file/s and $\mu=0.564$ files/s. It can be seen from Fig. 4 that both $\eta$ and $D$ monotonously increases with $\tau$ no matter what $f$ is, which means the offloading efficiency is improved at the expense of the mean delay. However, it is very interesting to see that, in the whole range of $\tau$, the increments of $\eta$ are the same while those of $D$ are different under two application scenarios. For example, when the increment of $D$ is $58$s when $f=0.005s^{-1}$, and that is $242$s when $f=0.025s^{-1}$, though the increments of $\eta$ in both cases are 0.7. This implies that the terminal with a higher mobility or a larger $f$ tends to pay a smaller cost in the mean delay to obtain the same increment of the offloading efficiency, which is visualized by Fig. 5 where $D$ is plotted as a function of $\eta$.
\begin{figure}[btp]
\centering
\subfigure[efficiency vs. deadline]{\includegraphics[width=0.3\textwidth]{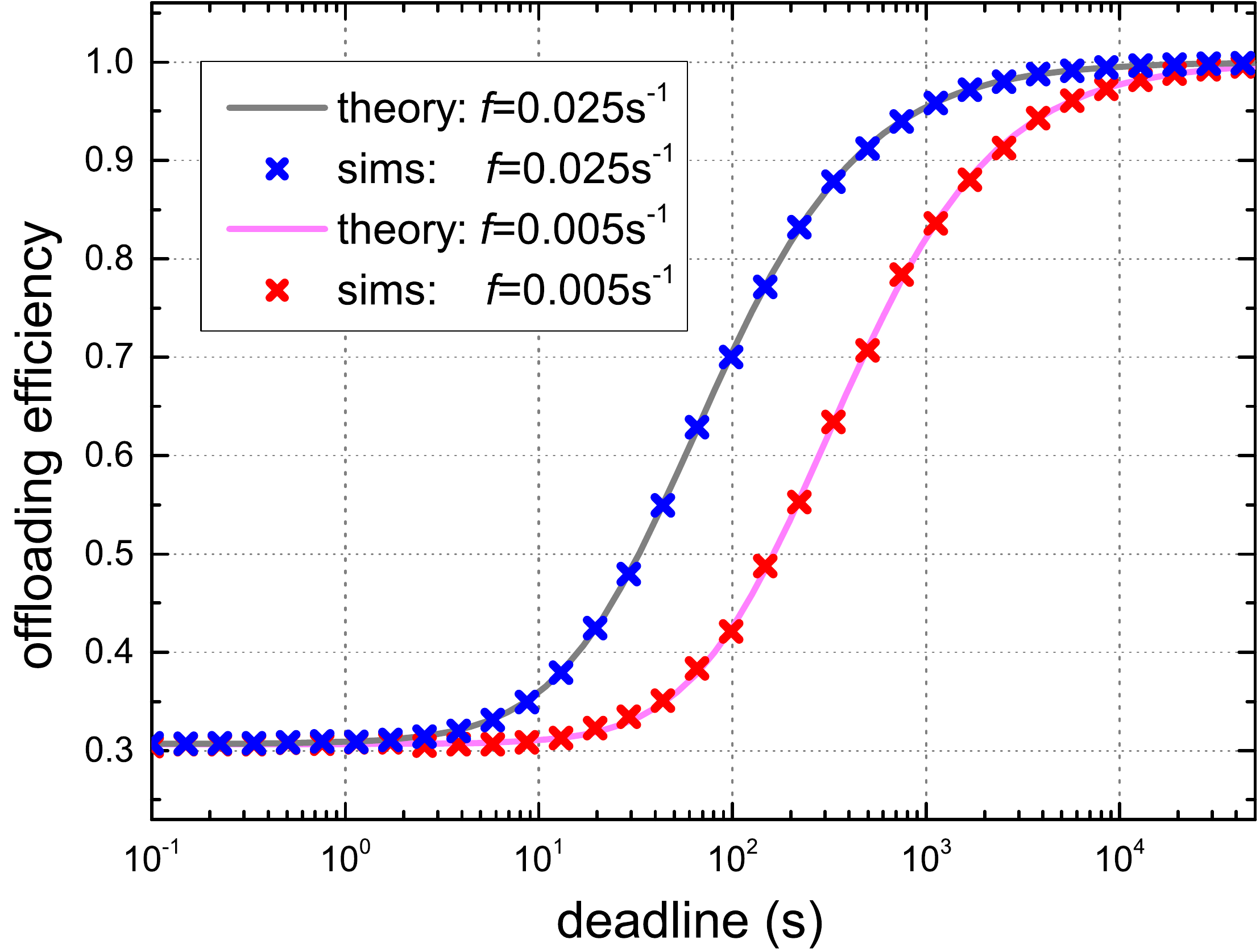}}
\subfigure[delay vs. deadline]{\includegraphics[width=0.3\textwidth]{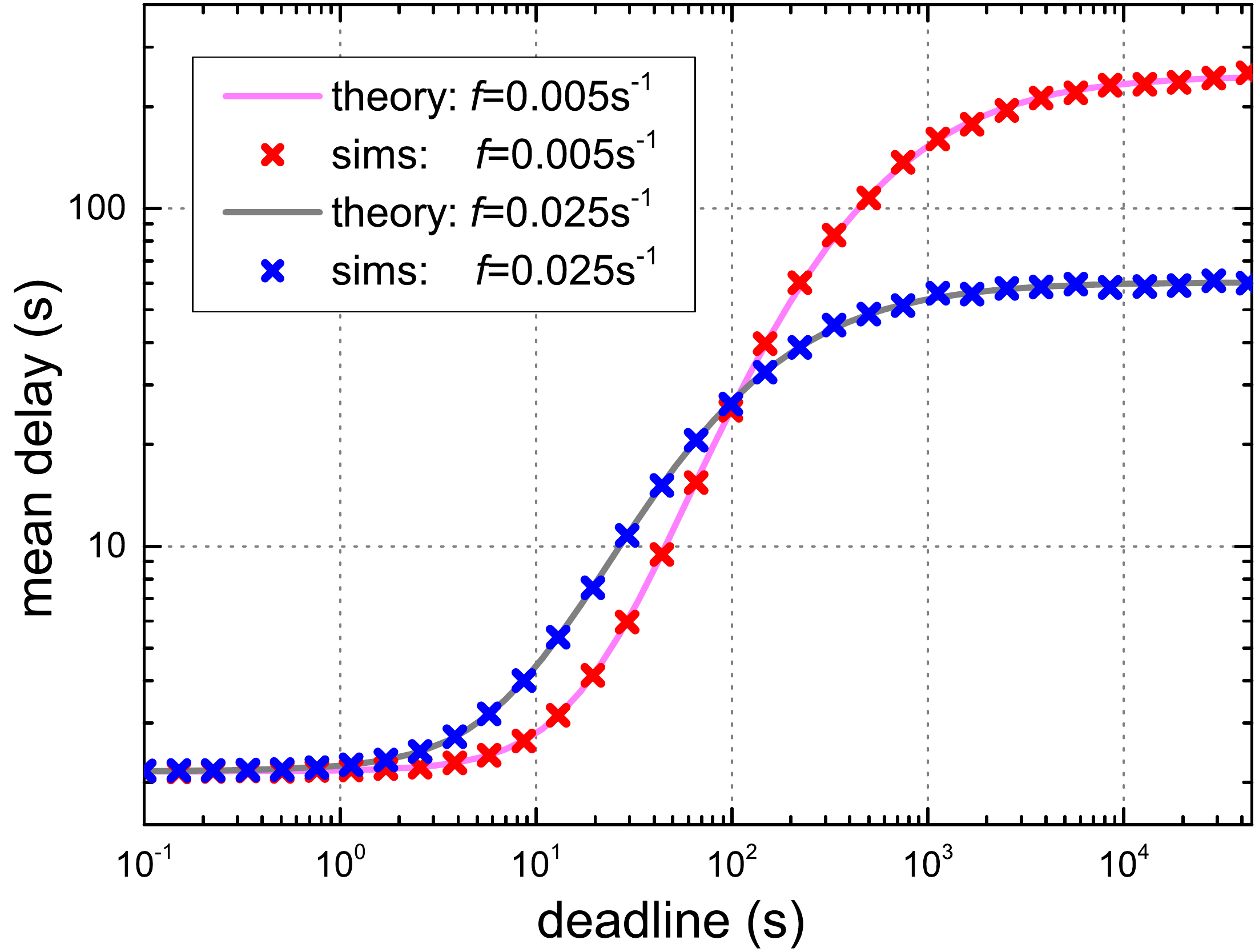}}
\caption{Performance of the delayed offloading when $\mu_1=\mu_2$}
\end{figure}

\begin{figure}[b]
\centering
\includegraphics[width=0.3\textwidth]{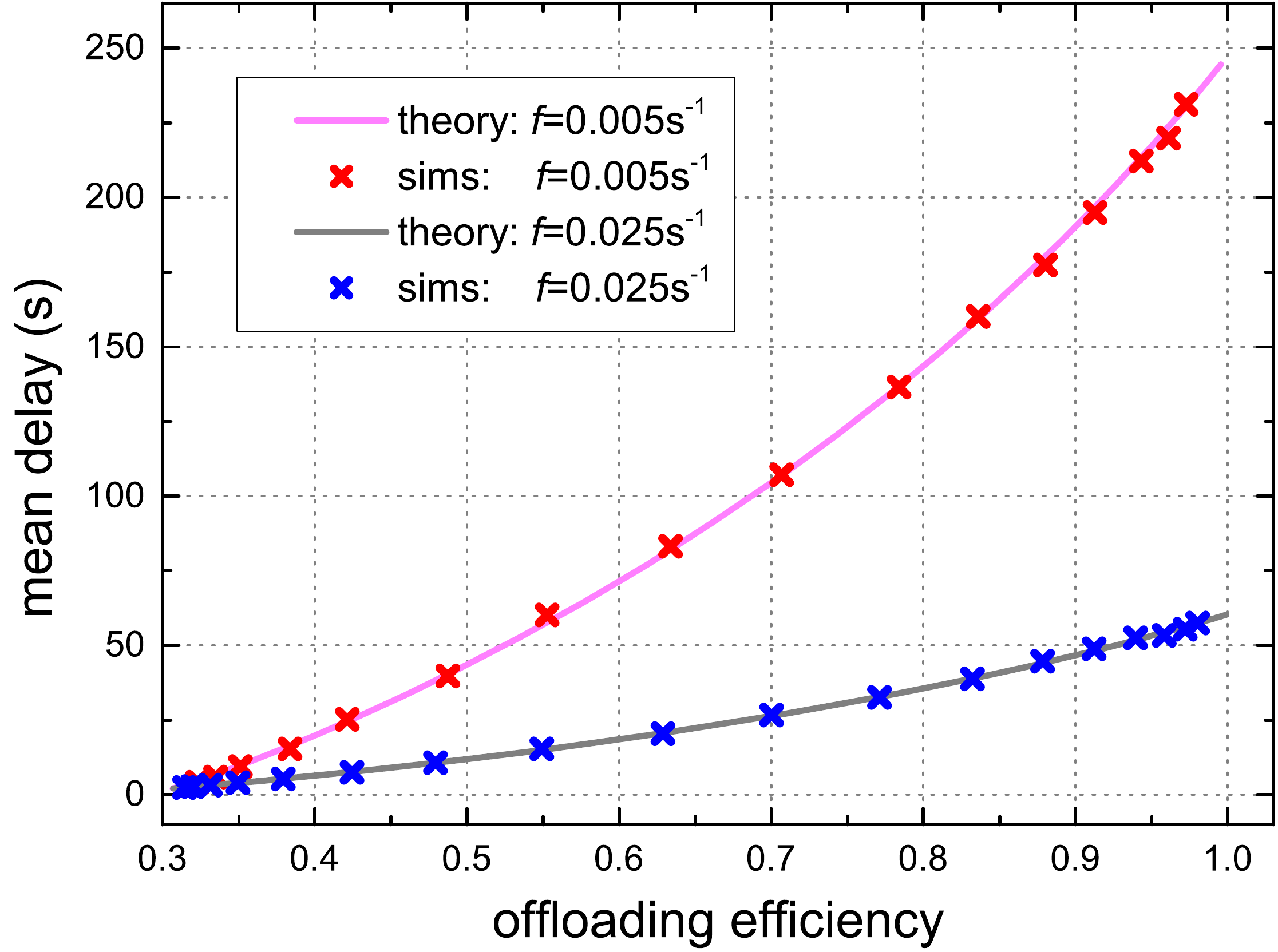}
\caption{Delay vs. efficiency when $\mu_1=\mu_2$.}
\end{figure}\par

Based on the analytical results in Section III, we explain this point by considering two extreme cases of $\tau$ as follows. When $\tau$ is very small, the delayed service state in Fig. 2 disappears and the terminal will transmit the file immediately after it losses the Wi-Fi signal. In this case, though there are Wi-Fi state and cellular state as well as the state transitions, the transmission rate $\mu$ keeps unchanged over time, which implies that the service process and thus the queue length is independent of the service state transitions. In other words, the \emph{M/MMSP/1} queue now reduces to an \emph{M/M/1} queue with service rate $\mu$. It follows that $p_{0,2}=\pi_2 p_0=\pi_2 \left(1-\lambda/\mu\right)$, and thus $\eta$ in (17) is now equal to $\pi_2$, no matter what $f$ is. Also, the mean delay in (16) changes to $D=1/\left(\mu-\lambda\right)$ for all $f$s. On the other hand, when $\tau$ is extremely large, the cellular state in Fig. 2 disappears, and $\pi_1=\hat{\pi}_1=0$ accordingly. In this case, the offloading process is actually an \emph{M/MMSP/1} with two service states, Wi-Fi state and delayed state, and the terminal only transmits traffic via Wi-Fi. Thus, $\eta$ in (17) increases to 1, for all $f$s. Moreover, it is easy to show that $D$ in (16) goes to
\begin{equation}
D^*=\frac{1}{R\mu-\lambda}\left[1+\frac{R\left(1-R\right)^2\mu}{f}\right],
\end{equation}
if $\tau$ approaches to infinity. It is obvious that $D^*$ decreases with the mobility $f$.
\begin{figure}[htbp]
\centering
\subfigure[$\mu_1<\mu_2$]{\includegraphics[width=0.3\textwidth]{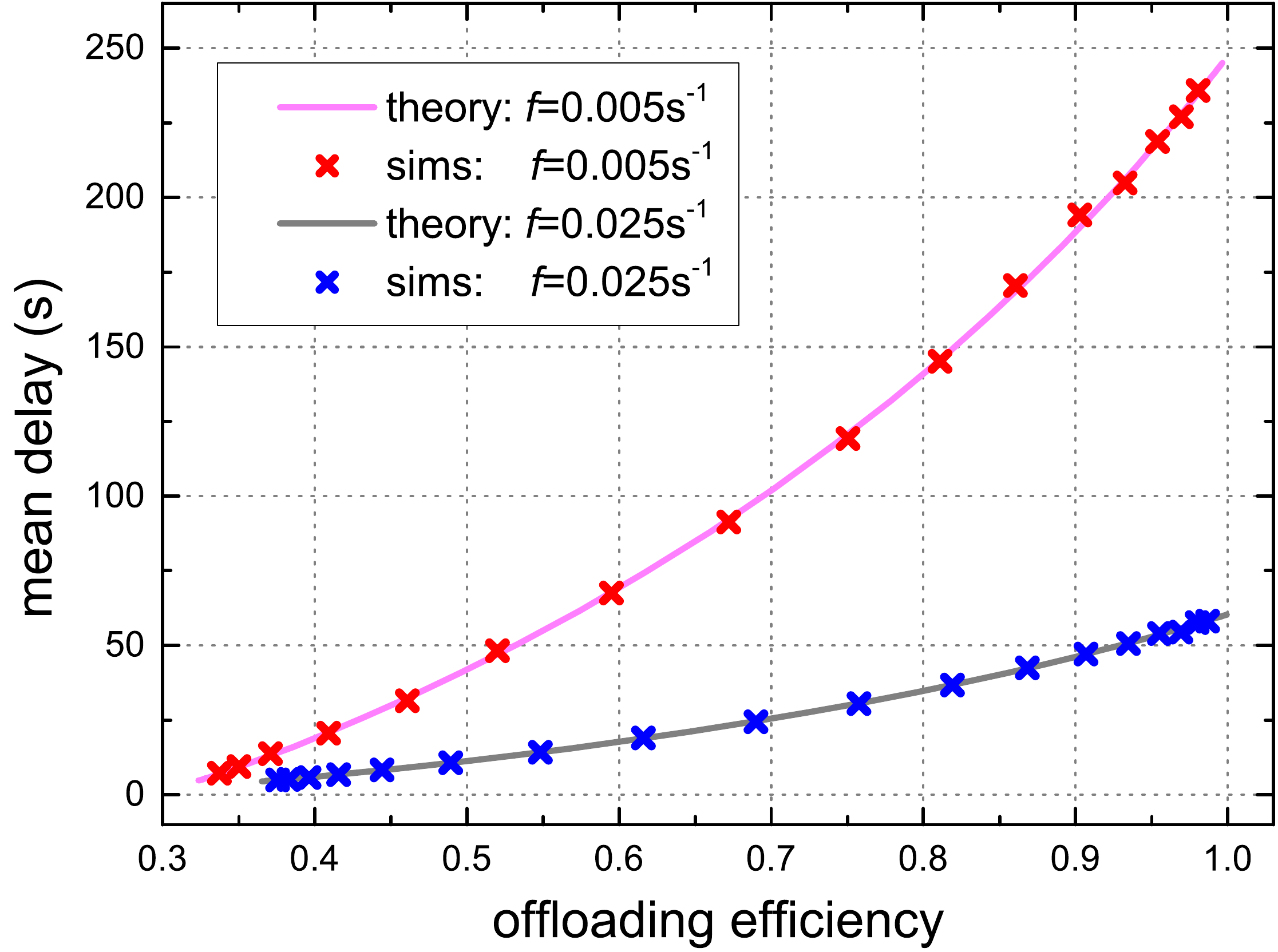}}
\subfigure[$\mu_1>\mu_2$]{\includegraphics[width=0.3\textwidth]{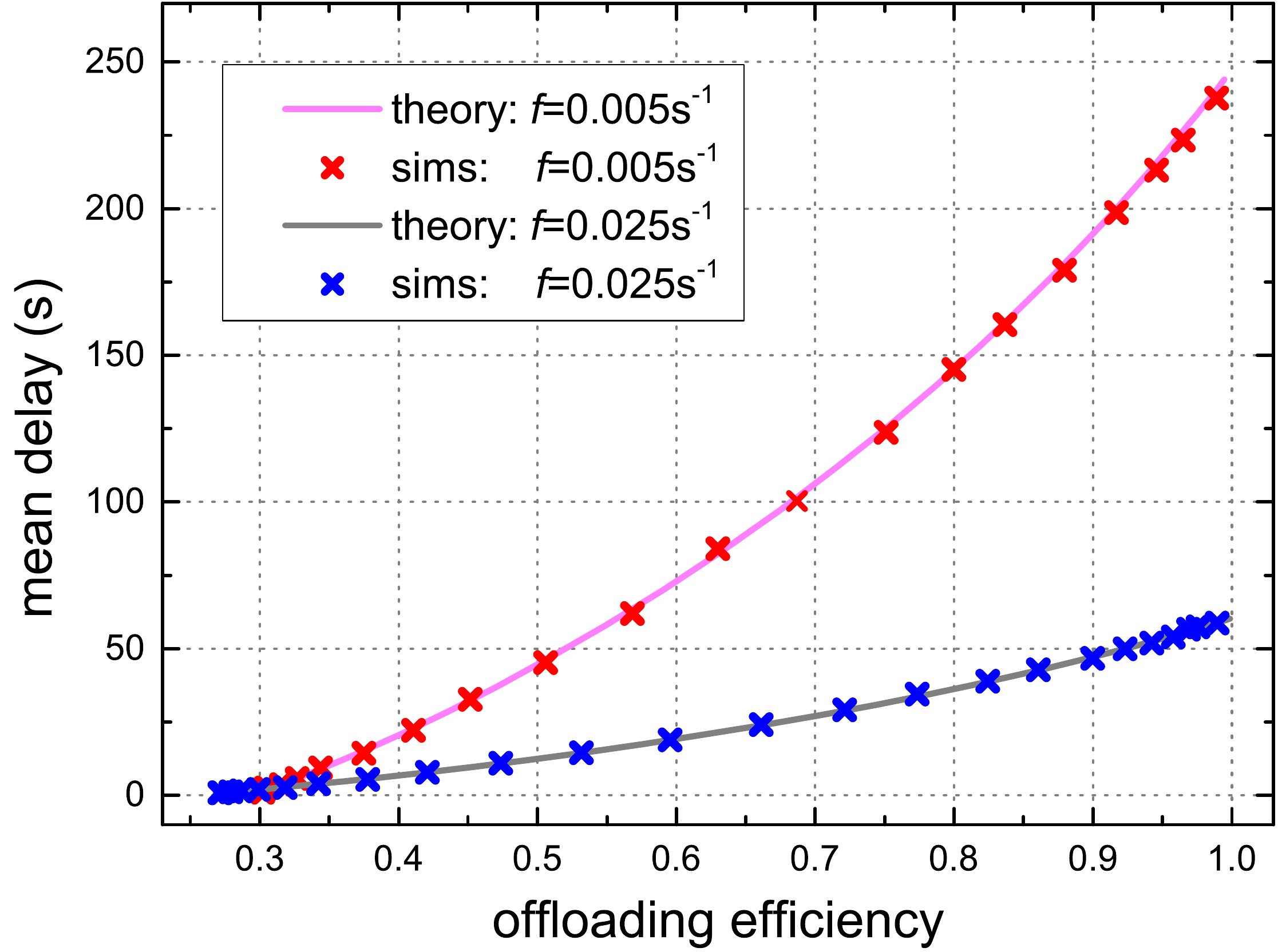}}
\caption{Delay vs. efficiency when (a) $\mu_1<\mu_2$ (b) $\mu_1>\mu_2$.}
\end{figure}

\subsection{$\mu_1\neq\mu_2$}
We also study the relationship between the mean delay and the efficiency when $\mu_1\neq\mu_2$ in Fig. 6, where all the parameters except $\mu_1$ and $\mu_2$ are all the same with those in Fig. 4. Especially, $\mu_1=0.6$ files/s and $\mu_2=1.28$ files/s in Fig. 6(a) and $\mu_1=10$ files/s and $\mu_2=1.28$ files/s in Fig. 6(b). The results in Fig. 6 show that the mobility of the terminal is helpful to reduce the expense of mean delay during the increase of offloading efficiency.

We thus conclude from the above discussion that the mobility of mobile users can reduce the delay incurred by the delay offloading. Based on this conclusion, mobile users can optimize the system performance according to their mobility. For example, a user has a delay requirement based on the delay tolerance. Our conclusion shows that the cost of delay to enhance the offloading efficiency is small when the mobility of the user is high. In this case, the user can greatly increase the deadline to improve the offloading efficiency while ensuring the delay requirement.

\section{Conclusion}
In this paper, we analyze the delayed offloading problem of currently commercial mobile terminals. We develop a three-state \emph{M/MMSP/1} queueing model to derive the offloading efficiency and the mean delay. Through the analysis, we find that the mobility of the users plays an important role in the tradeoff between the offloading efficiency and the mean delay. In particular, the mobility of mobile users can reduces the delay incurred by the delayed offloading.

\end{document}